%% file: IF07_topic5.tex
\documentclass[11pt]{article}
\usepackage{graphicx}
\usepackage[margin=1.25in]{geometry}
\usepackage[usenames,dvipsnames]{color}
\usepackage{url}
\usepackage[colorlinks = true,
            linkcolor = blue,
            urlcolor  = blue,
            citecolor = blue,
            anchorcolor = blue]{hyperref}

\newcommand\pubnumber{Transcendental Preprint }
\newcommand\pubdate{\today}

\def\Title#1{\begin{center} {\LARGE #1 } \end{center}}
\def\Author#1{\begin{center}{ \sc #1} \end{center}}

\newcommand\pubblock{\rightline{\begin{tabular}{l} \pubnumber\\
         \pubdate \end{tabular}}}
\newenvironment{Abstract}{\begin{quotation} \begin{center}
                       ABSTRACT
     \end{center}\bigskip  }{\end{quotation}}

\input workshopsymbols.tex

\newcommand\snowmass{\begin{center}\rule[-0.2in]{\hsize}{0.01in}\\\rule{\hsize}{0.01in}\\
\vskip 0.1in Submitted to the  Proceedings of the US Community Study\\ 
on the Future of Particle Physics (Snowmass 2021)\\ 
\rule{\hsize}{0.01in}\\\rule[+0.2in]{\hsize}{0.01in} \end{center}}

\begin{document}

\pubblock

\Title{Fast (optical) Links
}

\bigskip 

\Author{M.~Newcomer, J.~Ye, A.~Paramonov, M.~Garcia-Sciveres, A.~Prosser}


\medskip


\medskip

 \begin{Abstract}
\noindent In this write-up we focus on fast data links that read out particle detectors. This includes developments of ASICs, optical modules, identifying passive components such as fiber and connector, and a link system design. A review of the current status is provided. The goal of R\&D is to double (near term) and quadruple (longer term) the present data transmission rate for particle detectors, enabling a wide range of physics exploration and measurements.
\end{Abstract}

\snowmass

\section{Introduction and Current Status}

The past-decade has seen tremendous advances in particle detector developments, exemplified in pixel sensors~\cite{Hirono:2018rqw}, high-granularity calorimeters~\cite{Pitters:2018nsl} and time (of arrival) detectors~\cite{CERN-LHCC-2018-023, CMS:2667167} for MIPs in a jet. R\&D efforts are also seen in dedicated sensor readout ASICs~\cite{TWEPP2019:ALTIROC, TWEPP2019:ETROC}, on-detector readout electronics and data transmission to off-detector processing units~\cite{CERN-LHCC-2017-018}. In data transmission we have progressed from tens of mega-bits per second (Mbps) per channel to the present 10 Gbps per channel, and migrated from mainly coaxial-cable based transmission medium to optical fibers. At the beginning of the LHC experiments optical link systems were constructed using custom components, COTS (commercial off the shelf) ICs (example: the G-Link chip-set~\cite{Liu:2011zza}) that are tested to be radiation tolerant, and ASICs developed by CERN (example: the GOL serializer~\cite{Moreira:588665}). The transmission speeds range from 80 Mbps to 1.6 Gbps per fiber. With the increase of requirements in data rate and radiation tolerance, and the limit on power dissipation, ASICs become the only choice in constructing the on-detector side of the optical links with a data rate about 5 Gbps~\cite{Moreira:1235836} per fiber for the Phase-I upgrades of the LHC detectors and 10 Gbps~\cite{Xiao:2016dvu} per fiber for the Phase-II upgrades for the HL-LHC. The optical modules also become fully customized~\cite{CERN:VL, Zhao:2016czy} for both the Phase-I and Phase-II constructions due to the special requirements on module dimension, channel density, and tolerance to radiation and magnetic field in particle detectors. The electrical-optical signal converters and fibers are identified in the CERN led common project Versatile Link~\cite{CERN:VL, Huffman_2014} and the focus is now on VCSEL (vertical-cavity surface-emitting laser) and multi-mode fibers that are characterized to be radiation tolerant for applications in the LHC experiments. Given the size of these detectors and the added transmission loss due to radiation a nominal fiber length is chosen to be 150 meters in these R\&D projects~\cite{CERN:Troska}, which matches well with the short-range (SR) fiber optics data transmission in industry, mostly for data centers. This way the R\&D in HEP can benefit the advances in industry. Without the requirements of radiation tolerance and very stringent low-power dissipation on components, the data rates in industry for SR fiber links have reached 56 Gbps thanks to fast IC technologies in 28 or even 14 nm CMOS, and transmission schemes such as PAM4 (pulse amplitude modulation with 4 levels)~\cite{9619473, Hwang:19}. Although in HEP we use COTS, especially FPGA chips for the off-detector electronics, we do not benefit on data bandwidth beyond 10 Gbps because the current on-detector transmitter operates up to 10 Gbps. This results in large number of fibers between the on-detector and off-detector electronics (for example the ATLAS LAr readout upgrade will increase the number of fibers from the current some 2,000 to about 40,000) and very low efficiency in the use of the input bandwidth of the off-detector electronics, potentially increasing the number of PCBs and crates there. This limitation may also compromise the detector development preventing its full potential in exploring new physics or providing high precision measurements. For future detectors in HEP, there is a need to increase the channel bandwidth of optical link in the on-detector electronics. The key to this improvement are the data transmitter ASIC and the optical modules. 

\par Electrical-to-optical conversion can be integrated into the data transmitter ASIC (fiber-to-chip) or it can be done by a separate optical module. The fiber-to-chip approach can be realized using Si-photonics technology~\cite{9380443}. The separate optical module can be built using Si-photonics, VCSEL, or other technologies. Si-photonic optical circuits can be fabricated in the same wafer together with the electrical circuits using a conventional CMOS process inside the data transmitted ASIC or as a separate optical module~\cite{doi:10.1063/5.0050117}. So far the majority of fiber-optical data transmission in HEP experiments are relying on VCSELs and EELs~\cite{Huffman_2014, Zeng_2017}. The Si-photonics technology is relatively new to HEP; it has been explored since 2011 for use in the present and future experiments~\cite{Drake_2014, Kraxner:2019Il}. Modern commercial laser-based transceivers operate at 50 Gbps per optical link in 400G QSFP modules with power consumption of about 30 pJ/bit~\cite{Finisar_transceivers}. Comparable 400G QSFP Si-pho transceivers from Intel operate at 100 Gbps per fiber and consume about 20\% less power~\cite{Intel_transceivers}. Intel hopes to scale its silicon photonics platform up to 1 Tb/s per fiber at 1pJ of energy consumed per bit, reaching distances of up to 1 km. Both technologies offer sufficient link speeds for the future HEP experiments so their performance in radiation and cryogenic environments is the deciding factor. 

\par It is beneficial to develop higher speed serial data links to reduce the cost of the detector readout system. Higher bandwidth per a fiber-optical link decreases the total number of fibers. The FPGA-based off-detector readout system also gets cheaper when the links operate at high data rates. The bandwidth per link can be increased electronically (e.g. by serializing data faster)  and by using Wavelength Division Multiplexing (WDM).

\section{Proposed R\&D}

CERN through the common project lpGBT~\cite{TWEPP2019:lpGBT, CERN:lpGBT} has demonstrated 10.24 Gbps transmission speed in the first prototype SerDes (serializer-deserializer) ASIC, also called lpGBT. This ASIC, with its custom transmission protocol and forward error correction capability, is being qualified for applications in the LHC experiments and will be widely used in the upgrades for the HL-LHC or even beyond. CERN is forming working groups to study R\&D topics for future experiments~\cite{CERN:futureRD}. Detector data transmission (WG6) is among the topics and a goal of 56 Gbps per fiber using 28 nm CMOS technology and PAM4 is in discussion. In the US there has been R\&D work to build on the lpGBT design with a PAM4 scheme to reach 20.48 Gbps per fiber ~\cite{Zhang_2022}. The test chip GBS20 uses two of the lpGBT serializer design blocks with the shared PLL (also from lpGBT) following the scheme in the ASIC LOCx2 in ATLAS LAr trigger upgrade~\cite{Xiao:2016dvu}. A complete optical transmitter GBT20 has also been prototyped to demonstrate the transmission of 20.48 Gbps in one fiber through VCSEL~\cite{Zhang_2022}. As a near term goal the R\&D on ASIC and optical transmitter is aimed to double the current transmission speed. On the transmitting side, the R\&D work to be carried out is the implementation of the full lpGBT transmission protocol, ASIC package using the same C4-BGA developed for lpGBT, and to finalize the development of the optical module GBT20 with a robust fiber attachment scheme. On the receiving side, as 20.48 Gbps is not an industrial standard in PAM4 transmission, also because one needs to maintain this data rate (20.48 Gbps NRZ) in the electrical interface to widen the selection of FPGA chips, the ASIC (GBD20) and the optical receiver module (GBR20) will need to be developed. This work has already generated interest in CERN, as this will also serve as a pilot project in WG6 and the R\&D projects defined there to pursue higher speeds with 28 nm CMOS and PAM4. So strategically in the near-term R\&Ds should be supported on ASICs and optical modules that double the current 10 Gbps transmission speed. These R\&Ds should be in collaborations with CERN to benefit from the design blocks and expertise of the common projects. In the long-term there should be support for hardware groups to collaborate in the CERN led R\&D projects that aim at higher than 20 Gbps transmission using more advanced technologies than the current 65nm. 

A parallel US effort aims at a different readout architecture optimized for tracking detectors. This is being pursued under SBIR by Freedom Photonics in collaboration with UC Santa Barbara, Fermilab, and LBNL. There are also discussions with the CERN effort mentioned above as there is commonality in radiation hardening of ring resonator modulators. This architecture is better thought of as a conveyor belt rather than a fire-hose. The conveyor belt incrementally adds data to one fiber along its path, while the fire-hose aggregates a large amount of data first and then sends it down the fiber. The fire-hose needs a high speed aggregator chip (a la LpGBT) and a protocol for combining data. The conveyor belt has no on-detector high speed chip and adds data from different front end chips to the same fiber using wavelength domain multiplexing (WDM). This well suited to a distributed data source such as a tracking detector. The WDM is implemented with silicon photonic ring resonators. Furthermore, all control is off-detector by taking advantage of tunable laser technology.     

\subsection{ASIC Developments}

\par Collaboration with CERN and in the HEP community on ASIC needs to be encouraged and supported. As CERN, through the lpGBT ASIC and its HEP specific transmission protocol, has established eco-system of HEP detector data transmission and communication (clocking, configuring, control and monitoring), future developments need to follow this system and take advantage of existing design blocks. For high speed data transmission off the detector, the emphasis should be placed on PAM4 while exploring other techniques. In this regard, the 65 nm based GBS20 serves as a pioneer and that should be continued. The receiver ASIC GBD20 also needs to be supported. These efforts based on 65 nm technology will not only provides components to a 20 Gbps per fiber system for near future application, they will also serve as learning steps towards ASICs using 28 nm technology aiming at 40 Gbps per fiber and beyond. 

\subsection{Optical Module Developments}

\par So far and by far the mainstream optical modules used in HEP are VTRx~\cite{CERN:VL} and MTRx~\cite{Zhao:2016czy}. The next generation of VTRx, called VTRx+~\cite{CERN:VL}, will be the only choice in the detector upgrades for the HL-LHC. These optical modules or transmitter/transceivers are all VCSEL based with a direct driver circuit, packaged with custom designs suitable for HEP specific applications. The VCSELs and photo diodes are commercially available parts pre-selected based on the radiation tolerance~\cite{CERN:VL, Zhao:2016czy}. This trend will continue. R\&D work has already been reported in TWEPP~\cite{Huang:2781874} in both LC (single channel) and MT (array optics) optical connector formats. While exploring other technology such as si-photonics is forward looking, before a concrete prototype for HEP application is demonstrated, the direct-driving VCSEL based optical module development needs to be the mainstream. 


\par The majority of the R\&D for radiation-tolerance Silicon photonics has been conducted by CERN and other European institutions while the US initiatives are behind.  CERN has evaluated radiation tolerance of Mach-Zehnder interferometers, ring resonators, and other components~\cite{Kraxner:2019Il,Prousalidi:2781875}. They observed good tolerance to displacement damage and found a method to reduce the effects of ionizing radiation. Evaluation of complete transceivers will be required once characterization of the individual components including the chip-to-fiber coupling is complete. The development of Si-photonics transceivers is a novel and challenging initiative for the global HEP community so close collaboration with commercial partners could speed up the development process and reduce the costs. The US ASIC community can readily benefit by acquiring experience in Si-photonics design given its experience with the 65 nm and newer CMOS processes (e.g. see 45SPCLO from Global Foundries or TSMC28~\cite{10.1117/12.2600958,8540508}). There is interest in using Si-photonics transceivers in cryogenic environment for quantum networks~\cite{10.1117/12.2546635} so there is possibility to use it also for low-temperature HEP experiments.  

\section{Conclusion}

\par Precision measurements in HEP with higher energy and/or higher luminosity motive detector R\&D. To read out future HEP detectors, fast (optical) links are needed with ever higher bandwidth, higher channel density, lower power consumption, and more importantly radiation tolerant to the HEP detector environments. Only years of strong R\&D work will lead to mature and reliable systems to be deployed. Detector data transmission was determined as a bottle-neck in HEP detector developments. With proper R\&D support the limitation from this bottle-neck will hopefully be resolved. 








\bibliographystyle{JHEP}
\bibliography{myreferences}  






\end{document}

%% file: workshopsymbols.tex
\def\beq{\begin{equation}}
\def\eeq#1{\label{#1}\end{equation}}
\def\eeqn{\end{equation}}

\newenvironment{Eqnarray}%
   {\arraycolsep 0.14em\begin{eqnarray}}{\end{eqnarray}}
\def\beqa{\begin{Eqnarray}}
\def\eeqa#1{\label{#1}\end{Eqnarray}}
\def\eeqan{\end{Eqnarray}}

\let\bar=\overbar

\def\lsim{\mathrel{\raise.3ex\hbox{$<$\kern-.75em\lower1ex\hbox{$\sim$}}}}
\def\gsim{\mathrel{\raise.3ex\hbox{$>$\kern-.75em\lower1ex\hbox{$\sim$}}}}

\def\del{\partial}
\def\Dslash{\not{\hbox{\kern-4pt $D$}}}
\def\dslash{\not{\hbox{\kern-2pt $\del$}}}
\def\pslash{\not{\hbox{\kern-2pt $p$}}}
\def\ETmiss{\not{\hbox{\kern-4pt $E$}}_T}

\def\Dlr{\mathrel{\raise1.5ex\hbox{$\leftrightarrow$\kern-1em\lower1.5ex\hbox{$D$}}}}

\def\MSB{{\bar{M \kern -2pt S}}}
\def\msb{{\bar{\scriptsize M \kern -1pt S}}}

\def\drb{{\bar{\scriptsize D \kern -1pt R}}}

